\PassOptionsToPackage{numbers,sort&compress}{natbib}

\documentclass[twocolumn,superscriptaddress,showpacs,longbibliography]{revtex4} 

\usepackage{graphicx}
\usepackage{color}
\usepackage{amsmath}
\usepackage{amssymb}
\usepackage{bm}
\usepackage{enumerate}
\usepackage{float}
\usepackage{comment}
\usepackage{svg}

\usepackage[colorlinks=true,allcolors=blue]{hyperref}

\DeclareMathAlphabet{\mathsfit}{T1}{\sfdefault}{\mddefault}{\sldefault}
\newif\ifdraft 

\begin{document}
\title{Engineering Magnetic Anisotropy in Permalloy Films via Atomic Force Nanolithography}

\author{Abhishek Naik}
\affiliation{Experimental Physics of Nanostructured Materials,  Department of Physics, Universit\'{e} de Li\`{e}ge, B-4000 Sart Tilman, Belgium}

\author{Cyril Delforge}
\affiliation{Experimental Physics of Nanostructured Materials,  Department of Physics, Universit\'{e} de Li\`{e}ge, B-4000 Sart Tilman, Belgium}

\author{Nicolas Lejeune}
\affiliation{Experimental Physics of Nanostructured Materials,  Department of Physics, Universit\'{e} de Li\`{e}ge, B-4000 Sart Tilman, Belgium}

\author{Daniel Stoffels}
\affiliation{Experimental Physics of Nanostructured Materials,  Department of Physics, Universit\'{e} de Li\`{e}ge, B-4000 Sart Tilman, Belgium}
\author{Joris Van de Vondel}
\affiliation{Quantum Solid-State Physics, Department of Physics and Astronomy, KU Leuven, Celestijnenlaan 200D, Leuven, B-3001, Belgium}

\author{Kristiaan Temst}
\affiliation{Quantum Solid-State Physics, Department of Physics and Astronomy, KU Leuven, Celestijnenlaan 200D, Leuven, B-3001, Belgium}
\affiliation{Imec, Kapeldreef 75, 3001 Leuven, Belgium}

\author{Alejandro V. Silhanek} \email{asilhanek@uliege.be}
\affiliation{Experimental Physics of Nanostructured Materials,  Department of Physics, Universit\'{e} de Li\`{e}ge, B-4000 Sart Tilman, Belgium}

\author{Emile Fourneau} 
\affiliation{Experimental Physics of Nanostructured Materials,  Department of Physics, Universit\'{e} de Li\`{e}ge, B-4000 Sart Tilman, Belgium}

\date{\today}

\begin{abstract}
Atomic force nanolithography provides a precise method for sculpting magnetic thin films, enabling controlled engineering of magnetic anisotropy in soft ferromagnets at the microscale. We demonstrate that nanoscale groove arrays patterned into permalloy (Ni$_{80}$Fe$_{20}$) films induce a robust in-plane uniaxial anisotropy, with the easy axis aligned along the groove direction. The anisotropy field is shown to increase with decreasing groove period and increasing engraving depth, offering continuous tunability of magnetic hardness within a single fabrication step. Artificially engraved microstructures further allow domain configurations and domain-wall trajectories to be directed along predefined pathways, exemplified by the creation of a chessboard-like magnetic landscape. Owing to its adaptability to diverse ferromagnetic materials and arbitrary corrugation geometries, this approach provides a versatile platform for tailoring in-plane magnetic anisotropy. Concrete applications are demonstrated in the design of magnonic elements and anisotropic magnetoresistance sensors.
\medskip
\\
\noindent\textbf{Keywords:} SAGE, AFM nanolithography, uniaxial magnetic anisotropy, corrugation, permalloy
\end{abstract}

\maketitle

\section{Introduction}
Ferromagnetic (FM) thin films can be engineered to display a broad spectrum of magnetic properties beyond their as-deposited state. In particular, controlling both the intensity and direction of magnetic anisotropy is crucial for various applications, such as magnetic sensors \cite{Lopez2018}, memory devices \cite{Ikeda2010} and magnetoresistive devices \cite{Zhan-2007,gijs_perpendicular_1995,djuzhev_magnetic-field_2015,barrera_tunable_2023}. Extensive research has focused on tailoring the magnetic anisotropy of soft FM materials by introducing uniaxial periodic geometrical modulations.

The response of periodically patterned thin films has been studied in the context of magnetoplasmonic crystals \cite{belyaev_permalloy-based_2019,belyaev_magnetic_2020}, high-frequency technologies, and magnon transport mechanisms \cite{korner_two-magnon_2013,sakharov_spin-wave_2019,montoncello_controlling_2021,izotov_tailoring_2022}. 
Notably, experimental studies have demonstrated that periodically engraved FM waveguides act as efficient magnonic crystals, opening band gaps in the magnon dispersion curve \cite{chumak_design_2009,frey_reflection-less_2020}. Furthermore, rippled FM waveguides enable spin-wave propagation in the absence of an external magnetic field while guiding waves along specific pathways \cite{turcan_spin_2021,klima_zero-field_2024}.

Various approaches to introduce corrugations have been explored across a wide range of magnetic materials and fabrication techniques. Among them, substrate-induced techniques dominate the literature, where the substrate is periodically patterned prior to FM thin-film deposition. In such cases, enhancements in magnetic anisotropy of up to two orders of magnitude have been reported for rippled FM films on large-scale patterned substrates, achieved through methods such as lithography and anisotropic etching \cite{oepts_enhanced_2000}, ion erosion \cite{chen_uniaxial_2012,fassbender_introducing_2009,bukharia_evolution_2020,koyiloth_vayalil_tailoring_2020,xu_micromorphology_2023}, and laser-induced periodic surface structuring \cite{arranz_form_2019,sanchez_anisotropy_2020,delgado-garcia_magnetization_2023}. 

Alternatively, an artificial uniaxial magnetic anisotropy (UMA) can also be induced by directly patterning the top surface of a FM film. This approach was first demonstrated in ultrathin Co and Fe films \cite{moroni_uniaxial_2003,bisio_isolating_2006} and later extended to thicker films \cite{arranz_limits_2014,pan_tuning_2018,langer_spin-wave_2019}.
The mechanism underlying artificial UMA is generally ascribed to the anisotropic dipolar fields generated by magnetic charges at the interfaces of the rippled film, which are well described by Schlömann's equation \cite{schlomann_demagnetizing_1970}. However, experimental observations indicate that the physics at play can vary depending on the amplitude, period, and shape of the periodic modulation, which are inherently linked to the fabrication method. Indeed, Schlömann’s model applies to sinusoidally-rippled FM films with a low undulation amplitude-to-thickness ratio, short pattern periods, and holds for films with in-plane (IP) and nearly uniform magnetization. Therefore, deviations arise in cases such as (i) when surface-induced magnetic anisotropy dominates over bulk-induced anisotropy in thin films \cite{chen_uniaxial_2012,bisio_isolating_2006,liedke_crossover_2013}, (ii) when domain configurations affect the magnetic behavior of thick films \cite{delgado-garcia_magnetization_2023}, or (iii) in the presence of asymmetric patterns \cite{izotov_tailoring_2022,xu_micromorphology_2023}.

\begin{figure*}[t]
    \centering
    \includegraphics[width=1.0\linewidth]{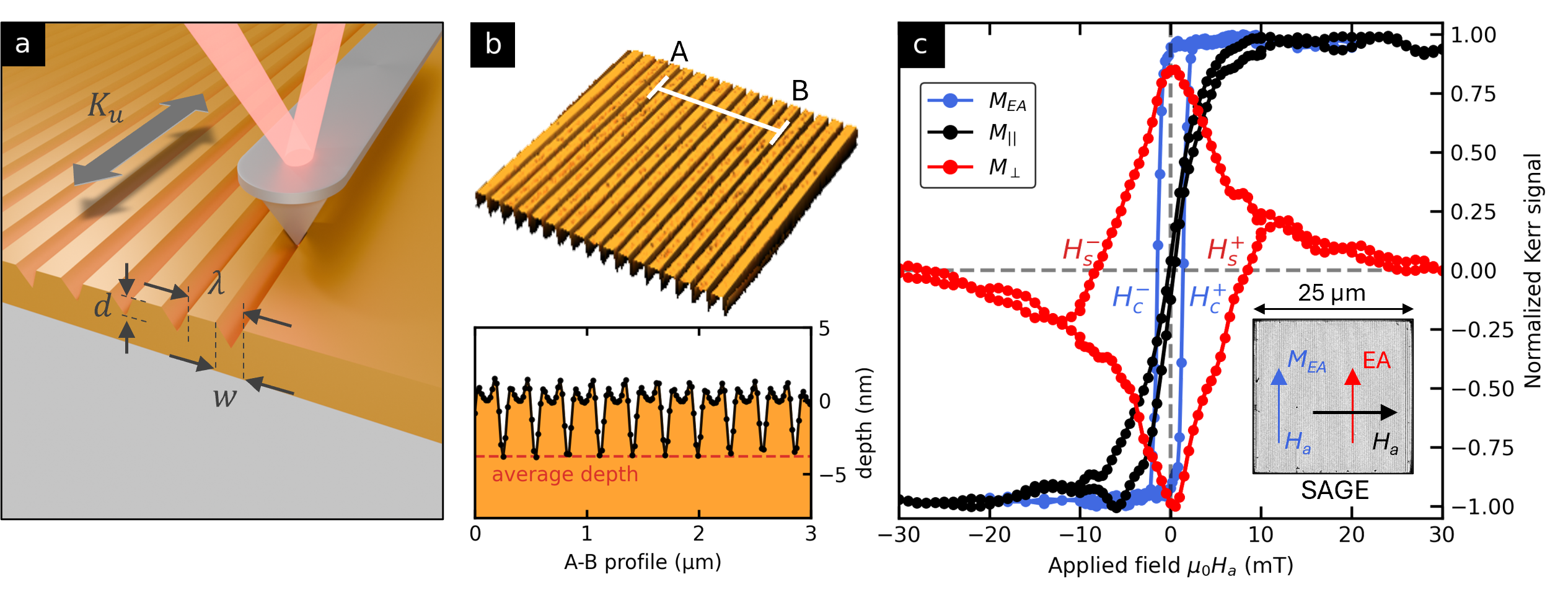}
    \caption{(a) Illustration of the surface artificial grooves engraving (SAGE) process, highlighting the relevant geometrical parameters. The grooves define the magnetic easy axis and the associated uniaxial magnetic anisotropy $K_u$ . (b) AFM topography image of a $5 \times 5$ $\mu$m$^2$ region with SAGE pattern. The cut-view profile along the 3 µm-long A-B section shows triangular grooves with homogeneous width and depth. (c) Normalized magnetic hysteresis loops showing the magnetization components parallel ($M_{\parallel}$, black dots) and perpendicular ($M_{\perp}$, red dots) to the magnetic field applied 5$^\circ$ off the SAGE hard axis (HA). The switching fields for the backward and forward branches are denoted as $H_s^-$ and $H_s^+$, respectively. The normalized hysteresis loop for the magnetization component along the SAGE easy axis ($M_{\mathrm{EA}}$, blue dots) is also shown for the field applied along the SAGE EA. The inset shows a scanning electron microscopy image of the surface structure of the investigated sample after SAGE. }
    \label{fig:fig1-sage}
\end{figure*}


Extensive research has explored artificial UMA in FM thin films with periodic surface modifications, spanning diverse fabrication techniques and deformation regimes (amplitude and period relative to film thickness). Most studies have focused on large-area structures with sinusoidal undulations, typically patterning chips of several squared millimeters. However, the use of surface engraving to induce UMA in micrometer-scale devices remains largely unexplored. This limitation arises from the greater sensitivity of small-scale structures to patterning quality: while large samples tolerate heterogeneities in undulation alignment, width, or depth due to averaging over thousands of grooves, micrometer-scale structures contain only a few tens of grooves, making fabrication imperfections much more detrimental. Therefore, achieving reliable UMA at these dimensions requires high-resolution nanolithography to minimize discrepancies between successive grooves. To the best of our knowledge, such local tunability of UMA has not yet been demonstrated. It remains an open question whether surface patterning can reliably induce UMA at reduced dimensions or be exploited to engineer non-uniform anisotropy textures. Overcoming these challenges could enable new opportunities in magnetic metamaterials \cite{fourneau_microscale_2023,lejeune_dimensional_2024,barrera_-chip_2025}, particularly for spin-wave optics, where waveguides with spatially varying magnetic properties are essential \cite{kiechle_spin-wave_2023}. 

In this work, we explore an alternative strategy for inducing uniaxial anisotropy in soft ferromagnetic thin films by engraving shallow artificial grooves using atomic-force nanolithography. This technique utilizes an Atomic Force Microscope (AFM) in contact mode, where the applied tip force is deliberately set to a high value to indent and reshape the surface. 
Through controlled nanoscale tip movement, this approach enables precise surface modification at the nanometer level \cite{delforge_investigation_2024,fang_nanometric_2022}. In contrast to conventional etching techniques, AFM nanolithography offers several advantages: it is compatible with a wide range of materials, operates under ambient conditions (without requiring vacuum, specific temperature, or humidity control), and enables low-cost, low-damage surface modification \cite{zhang_investigation_2020}. We demonstrate that this technique can be exploited to design non-uniform magnetic anisotropy landscapes capable of controlling the pinning and motion of specific magnetic textures  \cite{Albisetti2016}, such as domain walls (DWs) and magnetic vortices, by locally tuning both the intensity and direction of the uniaxial magnetic anisotropy. Finally, we establish the technological relevance of this approach by integrating it into magnetic-based devices, including sensitivity-enhanced anisotropic magnetoresistance (AMR) sensors and spin wave transport in the absence of an external field.

 

\begin{figure*}
    \centering
    \includegraphics[width=\linewidth]{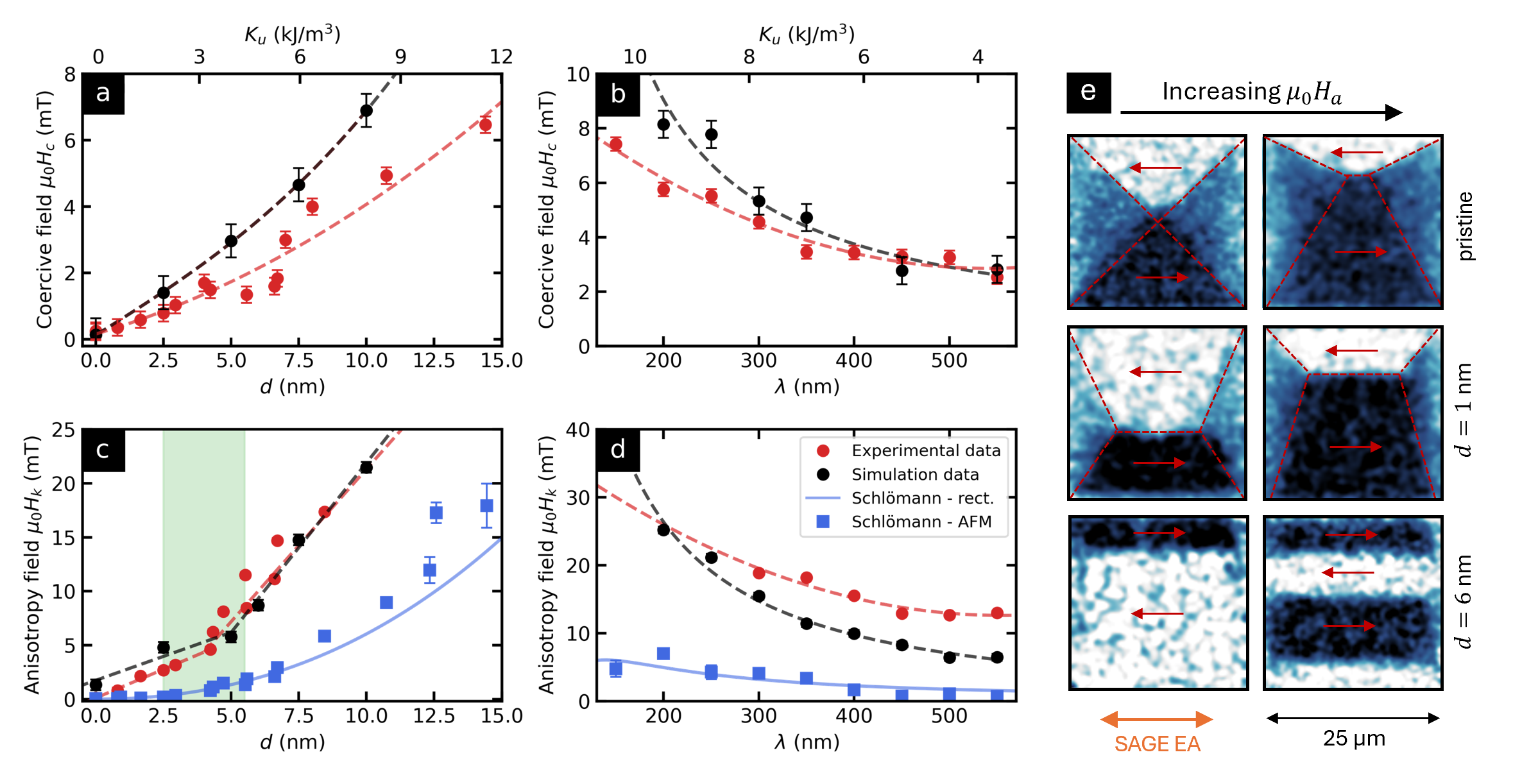}
    \caption{(a,b) Variation of the coercive field $H_c$ and (c,d) the anisotropy field $H_k$ as a function of the SAGE groove depth and period, respectively. Red dots represent experimental results, while black dots correspond to micromagnetic simulations incorporating a corrugated top surface. Dashed lines serve as guides to the eye. The upper horizontal axis in (a,b) shows the estimated energy density $K_u$, derived from micromagnetic simulations assuming an ideal homogeneous UMA. Blue solid lines represent the predictions of Schlömann’s model for ideal periodic rectangular grooves, while blue square symbols correspond to the same model using the roughness values measured by AFM. The green shaded area in (c) indicates the transition in the magnetization-reversal mechanism. (e) Magneto-Optic Kerr microscopy (MOKE) images of three devices with different scratching depths, acquired at two successive applied fields close to $H_a \simeq H_c$.}
    \label{fig:fig2-dnp}
\end{figure*}

\begin{table*}
\centering
\begin{tabular}{|l|c|c|c|c|}
\hline
\textbf{Technique} & \textbf{Amplitude (nm)} & \textbf{Period (nm)} & \textbf{$K_u$ (kJ/m$^3$)} & \textbf{Reference} \\
\hline
Laser interference photolithography and RIE
on substrate  & 180 & 250 & 7.1 & \cite{delgado-garcia_magnetization_2023} \\
Ion beam erosion on substrate & 1--6 & 24 & 1.5 & \cite{koyiloth_vayalil_tailoring_2020} \\
LIPSS on substrate & 45 & 240 & 4 & \cite{sanchez_anisotropy_2020} \\
SAGE & 0-15 & 100-550 & 0-12 & This work \\
\hline
\end{tabular}
\caption{Comparison of uniaxial anisotropy energy density ($K_u$) for different surface patterning techniques applied to 30 nm Permalloy films.}
\label{tab:Ku_comparison}
\end{table*}

\section{Results and discussion}

The samples under investigation consist of 30 nm-thick Permalloy (Py) deposited by electron beam evaporation on a silicon substrate with a 300 nm-thick SiO$_2$ layer. Prior to deposition, squares and disks of varying dimensions are patterned using electron-beam lithography and a conventional lift-off process. Subsequently, an IP uniaxial magnetic anisotropy is induced by AFM nanolitography using a diamond conical tip. The tip applies a constant force on the pristine sample surface, drawing an array of grooves with depth $d$, surface width $w$, and spacing period $\lambda$. The full process, referred to as \textit{Shallow Artificial Grooves Engraving} (SAGE), is illustrated in Fig. \ref{fig:fig1-sage}(a). The majority of residual material accumulated during the nanomachining process is swept away from the sample either during AFM imaging after the nanolithography process or via ultrasonic bath cleaning. SAGE average depth $d$ and width $w$ are then characterized using the AFM in tapping mode with a high-aspect-ratio tip, revealing a triangular shape for the grooves as shown in Fig. \ref{fig:fig1-sage}(b). Additional AFM measurements performed in different zones of the sample demonstrate that the SAGE depth, width, and spacing are constant over the entire sample, with small standard deviations of  0.2 nm, 2.5 nm, and 0.9 nm, respectively, for $d=3.7$ nm, $w=65.6$ nm, and $\lambda = 300$ nm. Both $d$ and $w$ can be tuned by adjusting either the applied force or the number of scans \cite{tseng_nanomachining_2010}, with both parameters being linearly correlated, as discussed in the Supplementary Material. However, deviations from this trend can be achieved by nanomachining multiple adjacent grooves that merge to form a wider groove without altering its depth \cite{zhang_investigation_2020}, an aspect not explored in this study.

The impact of the SAGE process on the magnetic behavior is characterized through magnetic hysteresis loops, obtained using a wide-field Magneto-Optical Kerr Effect microscope in longitudinal mode (L-MOKE). Further details on this technique can be found in the Methodology section. The SAGE-induced geometrical modifications in depth and periodicity significantly influence the intensity of the uniaxial magnetic anisotropy, thereby altering the shape of the hysteresis loop. This change is experimentally quantified by: (1) the coercive field $H_c$, defined as the mean of the coercive fields for the increasing ($H_c^{+}$) and decreasing ($H_c^{-}$) branches of the hysteresis loop measured with the field applied along the SAGE easy axis (EA); and (2) the anisotropy field $H_k$, defined as the mean of the switching fields ($H_s^{+}$ and $H_s^{-}$) extracted from the perpendicular magnetization component ($M_{\perp}$) hysteresis loops measured with the field applied at an angle of approximately 5$^\circ$ from the SAGE hard axis. This angle was chosen to maximize the signal and ensure abrupt switching at $H_s$, which closely approximates $H_k$ \cite{SANCHEZ2020167149,Perna,colino,Jimenez}.
The results of the MOKE measurements are shown in Fig.~\ref{fig:fig1-sage}(c), presenting the hysteresis loops of a $25 \times 25~\mu\text{m}^2$ Py square after the SAGE process. The data illustrate the evolution of the magnetization components parallel ($M_{\parallel}$, black dots) and perpendicular ($M_{\perp}$, red dots) to the applied magnetic field along the SAGE hard axis (HA), as well as the hysteresis loop measured with the field applied along the SAGE easy axis (EA), showing the corresponding magnetization component $M_{\mathrm{EA}}$ (blue dots).


\subsection{Effect of Groove Depth and Period on Magnetic Anisotropy}

For a given pristine structure (fixed material, design, and thickness), the intensity of SAGE-induced UMA can be controlled via the groove depth $d$ (coupled with the width $w$) and the periodicity of the pattern, defined by the wavelength $\lambda$. To investigate the range of UMA intensity achievable using SAGE and the impact of each geometrical parameter, a set of 25 $\times$ 25 µm$^2$ square devices with different SAGE depths and periods were fabricated. The main results are summarized in Fig. \ref{fig:fig2-dnp}. In the first batch of squares, SAGE was implemented with a constant periodicity of 300 nm while varying the corrugation depths from one square device to the next. As shown in Figure~\ref{fig:fig2-dnp}(a) and (c), L-MOKE measurements reveal that as the average groove depth $d$ increases, both the coercive field $\mu_0 H_c$ (panel (a), red dots) and the anisotropy field $\mu_0 H_k$ (panel (c), red dots) progressively increase. Coercive fields were measured for SAGE depths ranging from the pristine state up to around 14 nm, whereas $\mu_0 H_k$ were only determined for devices with depth up to approximately 8 nm. The anisotropy field of devices with grooves deeper than 8 nm are not reported, as domain saturation could not be achieved within the available magnetic field range, leading to unreliable determination of $H_s$. 
In panel (c), the increase of $H_k$ becomes steeper for devices with groove depths greater than approximately 4 nm. Interestingly, this transition corresponds to a change in the mechanism of magnetic domain reversal. This is apparent in the longitudinal MOKE images shown in panel (e), where squares with different SAGE depths (0, 1, and 5 nm, respectively) are displayed for two successive applied fields close to the coercive field. The red arrows indicate the magnetic domain direction. For the pristine sample and for the sample with $d=1$ nm SAGE, the magnetic contrast presents a Landau-like flux-closure pattern with a preferential direction along the SAGE pattern EA. An increase in the external field along the EA will induce a smooth displacement of the DWs. In contrast to that, for the square with SAGE 6-nm-deep, the complete reversal of the square occurs through successive abrupt nucleation of rectangular domains aligned with the SAGE main direction. It is worth noting that the nanowire-shaped domains suggest a strong pinning induced by the SAGE process, and typically involve a cluster of several grooves switching together. Moreover, the observed sequential magnetic flip, rather than the collective, abrupt nucleation and rotation of the full square, can be attributed to defects and irregularities in the SAGE depth and period. Optimizing the SAGE process (grooving and cleaning) could help achieve a fast and unison magnetic reversal process. An analysis of MOKE images and micromagnetic simulations obtained for each sample, as represented in panel (c), allows for the separation between these two regimes of magnetic reversal mechanisms. The transition zone, where distinguishing between these two regimes is not possible, is highlighted in the light green area in panel (c).

To support the experimental findings and validate the data analysis, we resort to micromagnetic simulations (see Methodology section) using two different approaches. In a first approach, the simulated geometry is a cuboid with rectangular grooves at the top surface with period, depth, and width mimicking the values measured by AFM. The simulation results (black dots) reported for the depth variation in panel (c) quantitatively match the MOKE measurement,  confirming that the model captures the essential magnetic switch mechanism of the real device. Interestingly, regarding Brown's paradox \cite{brown}, the slight quantitative difference between micromagnetic simulations and experiments suggests that the SAGE pattern topography and the micrometer-scale square geometry serve as sources of nucleation sites for magnetic domains. Indeed, the weak discrepancy is attributed to the impact of defects (such as grain boundaries, substrate roughness, intrinsic material anisotropy, etc.) not considered in the micromagnetic simulations. 
In a second approach, simulations have been performed considering perfect cuboids with flat topography, whereas the UMA induced by SAGE is mimicked by a homogeneously distributed anisotropy with density energy $K_{u}$. The correspondence between both approaches permits us to establish a relationship between $K_u$ (upper abscissa of Fig.~\ref{fig:fig2-dnp}(a) and (b)) with SAGE depth and SAGE period, (bottom abscissa of Fig.~\ref{fig:fig2-dnp}(a) and (b)) respectively, giving rise to a quantification of the anisotropy energy density induced by the SAGE process. Simulation results show an increase of $K_u$ proportional to $d$ and $\lambda$, with a slope of approximately 0.001 kJ/m$^3$/nm and -0.59 kJ/m$^3$/nm, respectively. The correspondence with the experimentally measured $H_c$ values in Fig \ref{fig:fig2-dnp}(a) suggests that anisotropy energies $K_u$ ranging from 0 to 8.0 kJ/m$^3$ are obtained in samples with SAGE depths of 0 to 10 nm. The same reasoning is done for the SAGE periodicity in the upper axis of panel (b). Those values have the same order of magnitude as results previously reported in the literature for UMA induced in Py films of comparable thickness ($\sim$30 nm) using other surface patterning techniques, such as laser-induced periodic surface structures (LIPSS) \cite{sanchez_anisotropy_2020}, ion irradiation \cite{koyiloth_vayalil_tailoring_2020}, or using laser interference photolithography with reactive ion etching \cite{delgado-garcia_magnetization_2023}, as summarized in Table \ref{tab:Ku_comparison}.

Finally, for the sake of completeness, the experimental measurements are compared with the prediction based on Schlömann's equation \cite{schlomann_demagnetizing_1970}, for which the theoretical roughness is calculated both considering perfect rectangular-shaped grooves (as designed in the micromagnetic simulations) and the measured sample roughness as determined by AFM topography. The resulting theoretical IP UMA field is given by
\begin{equation}
    H_{k}=\pi \dfrac{\sigma^2}{\lambda t}M_s=\pi\dfrac{d^2w(\lambda-w)}{\lambda^3t}M_s,
\end{equation}
with $M_s$ being the saturation magnetization, $\sigma$ the RMS roughness, and the last equality standing for the perfect rectangular grooves.

As shown in Fig.\ref{fig:fig2-dnp}(c) and (d), AFM-based results (blue square symbols) are in good agreement with the perfect rectangle model (blue line), supporting the choice of a rectangular profile of the grooves for the micromagnetic simulations (comparison with other geometries is done in the Supplementary Material). While the Schlömann's model shows an increase of the anisotropy field with the SAGE depth, as predicted by both the micromagnetic simulations and the experimental measurements, a clear quantitative discrepancy is observed. This discrepancy can be explained as follows: (i) for a deep SAGE pattern, the corrugation amplitude overcomes the critical thickness ($5d>t$ for Py) suggested by Liedke et al.\cite{liedke_crossover_2013}, below which the magnetization vector tends to follow the surface undulation instead of generating the dipolar charges as hypothesized by Schlömann's model. (ii) For shallow SAGE samples, the slight difference with experimental data is attributed to the influence of the sample's shape anisotropy, in which the flux-closure pattern formed by the magnetic domains pins the magnetization strongly along the square edges, leading to a higher value of $H_k$.

The effect of varying the period $\lambda$ is shown in Fig.\ref{fig:fig2-dnp}(b) and (d). Samples have been patterned by SAGE with a depth $d=8\pm 1$ nm (corresponding to $w=120$ nm) and a period in the range from $\lambda = 150$~nm to $\lambda = 550$~nm. MOKE results reveal an inverse dependence of the coercive field $H_c$ on the groove periodicity. A similar inverse trend is observed for the anisotropy field $H_k$ within the periodicity range of 250–550 nm. Periodicities below 250 nm are not reported here, for the same reason as explained previously for deep SAGE devices. The micromagnetic simulations show a good correlation with the experimental data, with slightly larger coercive fields obtained with the micromagnetic simulations, as observed in panel (a). However, it is worth noting that the common experimental value ($\lambda=300$ nm, $d=8$ nm deep) reported in Fig.\ref{fig:fig2-dnp}(c) do not exactly match the one reported in panel (b), respectively $\mu_0 H_k=20$ mT and $\mu_0 H_k=16$ mT. The difference is attributed to the intrinsic properties of the Py film, as samples were not fabricated during the same deposition session. The origin of this difference is explained in detail in the following section.

Overall, it is apparent that varying the depth and period of the grooves allows for deterministic control of the coercive field and anisotropy field of FM thin films, ranging approximately from 0 to 8 mT and 0 to 25 mT, respectively. Larger values of $H_k$ are expected for devices with SAGE depth greater than 10 nm and period below 150 nm. The experimental trend reported in Fig. \ref{fig:fig2-dnp}, along with additional micromagnetic simulations, suggests the possibility of achieving anisotropy fields exceeding 50 mT.

\subsection{Angular Dependence of In-Plane Magnetization and Coercive Field}

\begin{figure}
    \centering
    \includegraphics[width=0.9\linewidth]{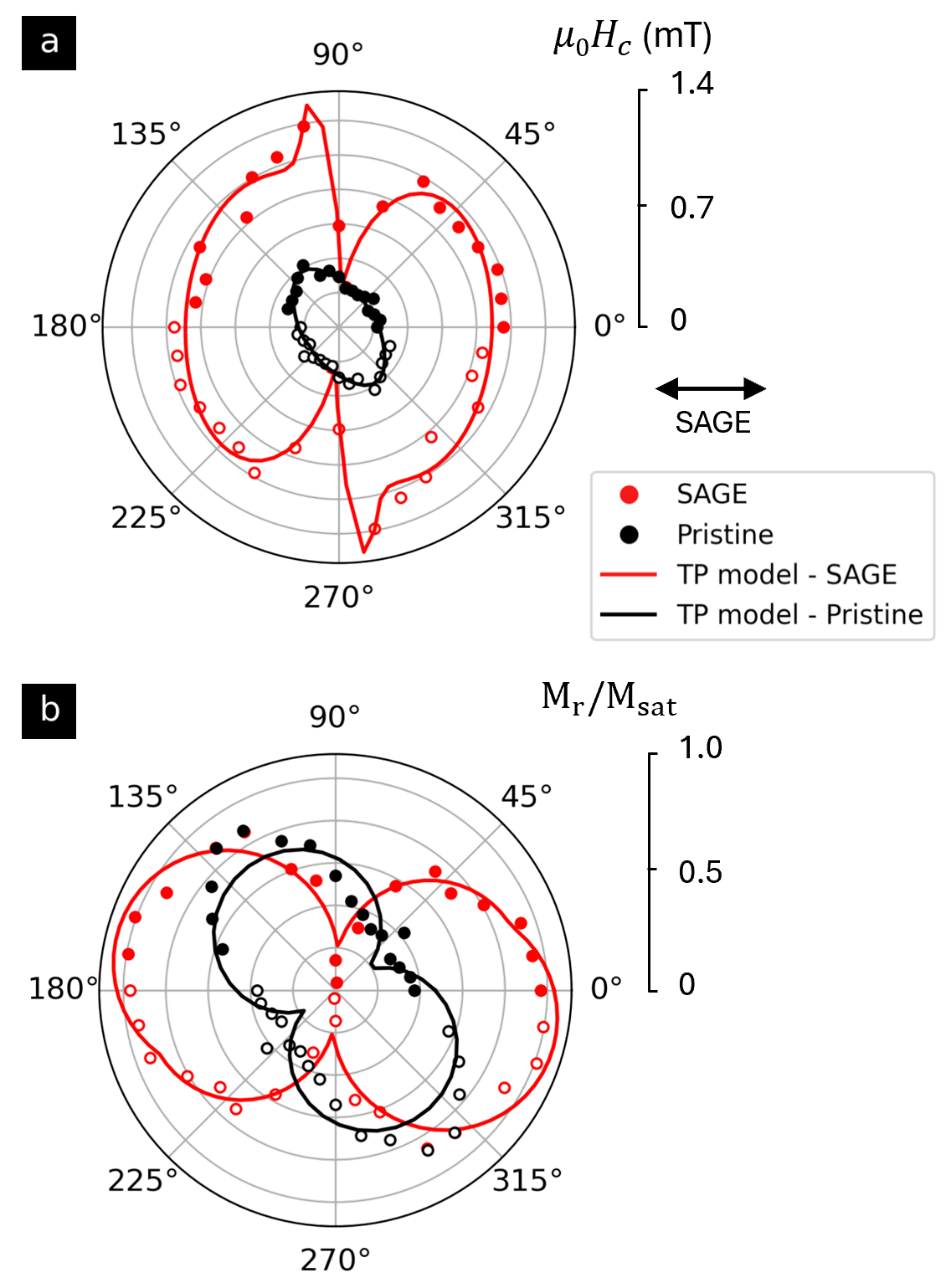}
    \caption{Angular dependence of (a) $H_c$ and (b) $M_r$ for SAGE-patterned ($d=1.5$ nm, $\lambda=600$ nm) and pristine 50-µm-wide Py disks. Experimental results have been obtained by L-MOKE. Results are fitted using the Two-Phase Pinning model, assuming the presence of two independent uniaxial anisotropies. Solid dots denote measured data and empty dots show their mirrored counterparts for clarity.}
    \label{fig:fig3-polar}
\end{figure}

Building upon the ability to engineer UMA via SAGE depth and period, the angular dependence of the magnetization reversal associated with SAGE-induced anisotropy is investigated. For this purpose, a series of $30$ nm-thick Py disks with a diameter of $50$ $\mu$m were fabricated, followed by the nanomachining of 1.5 nm-deep SAGE with different patterning periods, from 300 to 600 nm. The hysteresis loop of each sample is measured using L-MOKE while varying the in-plane deviation angle $\theta$ of the applied magnetic field with respect to the EA orientation. The obtained angular dependence of the coercive field $H_c(\theta)$ and the magnetization at remanence $M_r(\theta)$ of a sample with a SAGE period of 600 nm are shown in Fig. \ref{fig:fig3-polar}(a) and (b), respectively. For each plot, the SAGE-patterned sample (red dots) is shown together with the angular dependence of the pristine sample (black dots). Measurements have been performed on half a revolution (filled dots) and, for the sake of clarity, mirror values (empty dots) are also displayed, as expected based on the device symmetry. 

A characteristic two-fold angular variation for both $H_c$ and $M_r$ is obtained, which is a well-established signature of magnetic structures presenting UMA. However, the deduced EA of the device slightly deviates (less than 10$^\circ$) from the SAGE pattern orientation. This effect has been systematically observed in samples fabricated under identical conditions, even with different SAGE periodicity, as reported in the Supplementary Material. This deviation can be attributed to the presence of a non-negligible UMA in the pristine samples, where an EA along the 120$^\circ$ direction is observed, as shown in red in panels (a) and (b). This uniaxial anisotropy of the unpatterned sample originates from a slight tilt in the deposition angle during e-beam evaporation, which breaks the in-plane symmetry and preferentially aligns the magnetic easy axis. Such slight deviations from the expected EA have been reported at times, but they are rarely examined in depth in the literature \cite{liedke_crossover_2013, delgado-garcia_magnetization_2023}. Other sources of uniaxial anisotropy, such as substrate corrugation \cite{Zhan,barrera_tunable_2023}, can be ruled out in the present case.
The L-MOKE data for the $H_c(\theta)$ values of the SAGE-patterned samples exhibit an M-shaped polar response for the applied field along different orientations with respect to the SAGE main direction. These results are well reproduced by the two-phase pinning (TP) model \cite{neel1960laws,hazra2020magneto} depicted with solid lines in Fig.~\ref{fig:fig3-polar}. The model, which combines the Stoner–Wohlfarth and Kondorsky approaches, suggests that the magnetization reversal process begins with domain-wall depinning and progressive motion governed by the Kondorsky mechanism, followed at higher fields by coherent rotation of the magnetization (as described by the Stoner–Wohlfarth model), until full saturation is achieved \cite{suponev_angular_1996,barwal}. Consequently, the TP model predicts an angular variation of the coercive field and remanent magnetization following the expressions:

\begin{align}
    & H_{c}(\theta)=H_{c}(0)\dfrac{\cos{(\theta)}}{(1/\gamma)\sin{^2(\theta)}+\cos{^2(\theta)}},\\
    & M_r(\theta)=M_s\cos{(\theta)},
\end{align}

\noindent where $H_c(0)$ is the coercive field at $\theta$ = 0 (along the SAGE main direction), $N_x$ and $N_y$ are the in-plane demagnetization factors and the parameter $\gamma=(N_x+H_k/M_s)/N_y \simeq 1+DH_k/tM_s$ quantifies the intensity of the UMA in a disk of thickness $t$ and diameter $D$. 
For the SAGE-patterned samples, a proper fitting requires the linear combination of two TP models, one with EA lying along the SAGE main direction and another for the pristine UMA. The extracted value for the $\gamma$ parameters corresponding to the SAGE UMA is 2.5, resulting in a value of $\mu_0 H_k=1$ mT. The same value of $H_k$ is obtained from mean of the switching fields ($H_s^{+}$ and $H_s^{-}$), and the coercive field along the EA is also $\mu_0 H_c\simeq1$ mT. These values are in agreement with the trends reported in Fig. \ref{fig:fig2-dnp} and support the hypothesis that the TP pinning model captures the magnetic reversal mechanism of the SAGE-patterned disk with shallow depth and large periodicity. 

Finally, it is worth noting that, the pristine UMA being involuntary, its direction can vary from one batch of sample to the other, thus explaining the slight differences between the $H_c$ and $H_k$ values observed for similarly SAGE-patterned devices ($\lambda=300$ nm and $d=8$ nm) shown in panel (a,c) and (b,d) of Fig.\ref{fig:fig2-dnp}.

\begin{figure}[h!]
    \centering
    \includegraphics[width=1.0\linewidth]{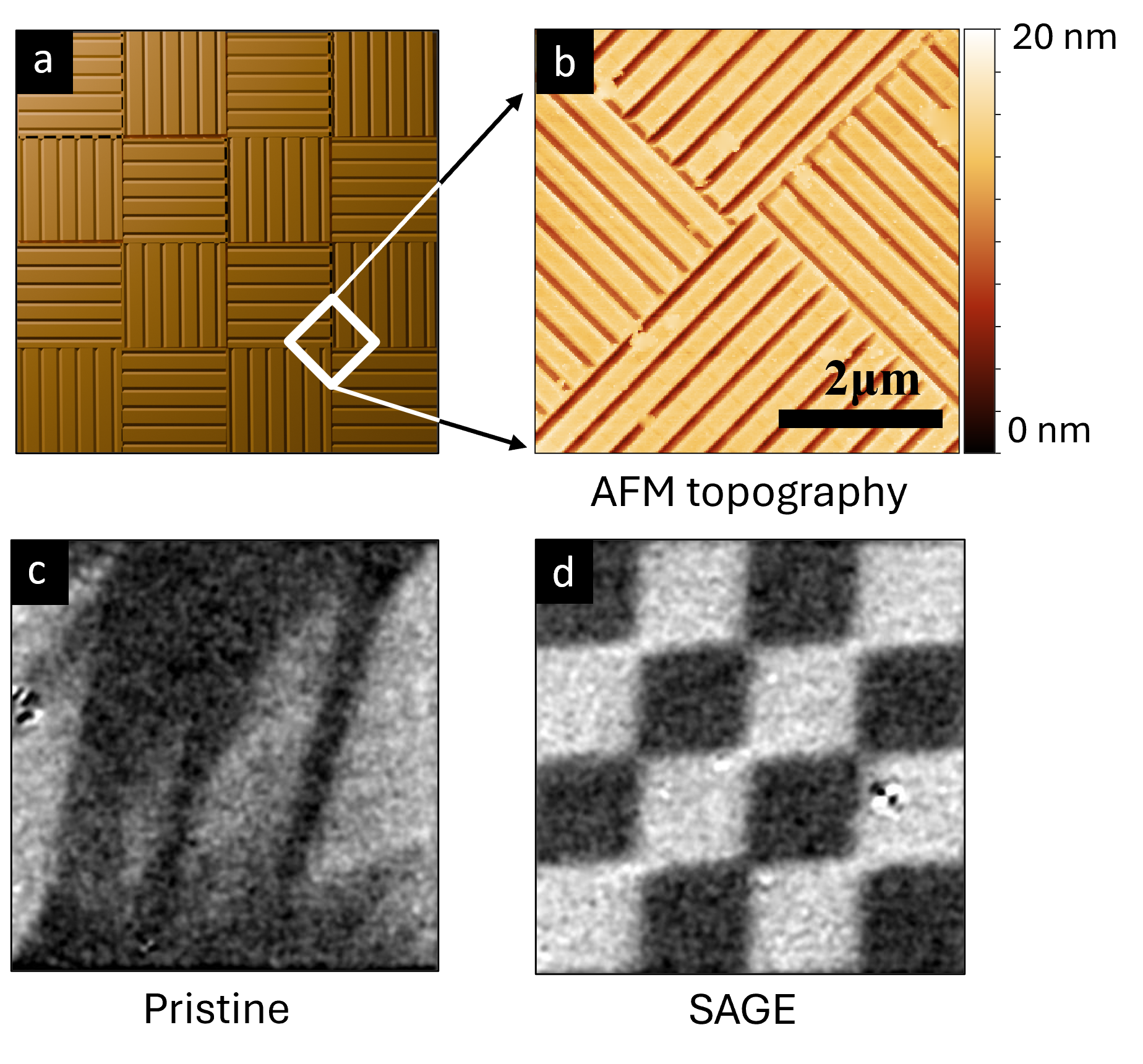}
    \caption{(a) Schematic illustration of the SAGE pattern designed to promote a chessboard-like domain arrangement. Each square has a side of 12.5 µm. (b) AFM topography at the junction between regions of alternating SAGE direction. (c–d) L-MOKE images showing (c) the domain configuration of a pristine square Py device, and (d) the emergence of a chessboard-like domain arrangement induced by the SAGE-patterned surface.}
    \label{fig:chess_moke}
\end{figure}

\subsection{Unconventional magnetic anisotropy landscape}

As discussed in the introduction, other techniques exist that permit the creation of artificial magnetic anisotropy on a large scale, such as tilted angle deposition \cite{ali2021anisotropic} or laser-induced periodic surface structuring \cite{arranz_form_2019}. However, a unique advantage of AFM nanolithography lies in its ability to alter the magnetic anisotropy locally, thereby allowing the fabrication of unconventional anisotropy landscapes. In order to illustrate this point, a 50~µm-wide chessboard-like pattern was fabricated using AFM-assisted SAGE. It consists of a periodic sequence of 12.5~µm-wide squares, each featuring 9.5~nm-deep grooves with a periodicity of $\lambda = 400$~nm and an EA oriented alternately along 0$^\circ$ and 90$^\circ$, as sketched in Fig.~\ref{fig:chess_moke}(a). Panel (b) shows an AFM topography at the junction of four neighboring patterned domains.  

As shown in Fig.~\ref{fig:chess_moke}(d), L-MOKE microscopy reveals that the patterned films develop deformed square-shaped magnetic monodomains arranged in a chessboard-like lattice in the remanent state after saturation by a magnetic field of 20 mT along the horizontal axis direction. This highly unnatural magnetic domain distribution demonstrates that the corrugation pattern acts as a robust template for texturing the local magnetic landscape with a resolution of a few micrometers. For the sake of comparison, Fig.~\ref{fig:chess_moke}(c) shows the domain configuration observed before introducing the periodic array of grooves. It is worth mentioning that the proposed approach complements the exchange bias-induced spatially modulated coercivity technique, which requires far more complex multilayer structures \cite{Buntinx}.

\subsection{Example of applications}

The ability to locally tailor the uniaxial anisotropy by performing SAGE on the surface of a soft FM film is showcased through two concrete applications. First, the feasibility of using SAGE to fabricate an anisotropic magnetoresistance (AMR) sensor is demonstrated, offering potential applications in magnetic field and current sensing, as well as in magnetic read heads. Second, the implementation of SAGE for designing a zero-field spin-wave (SW) waveguide is presented.

\subsubsection{Anisotropic magnetoresistance sensor without barberpole}

\begin{figure}
    \centering
    \includegraphics[width=0.8\linewidth]{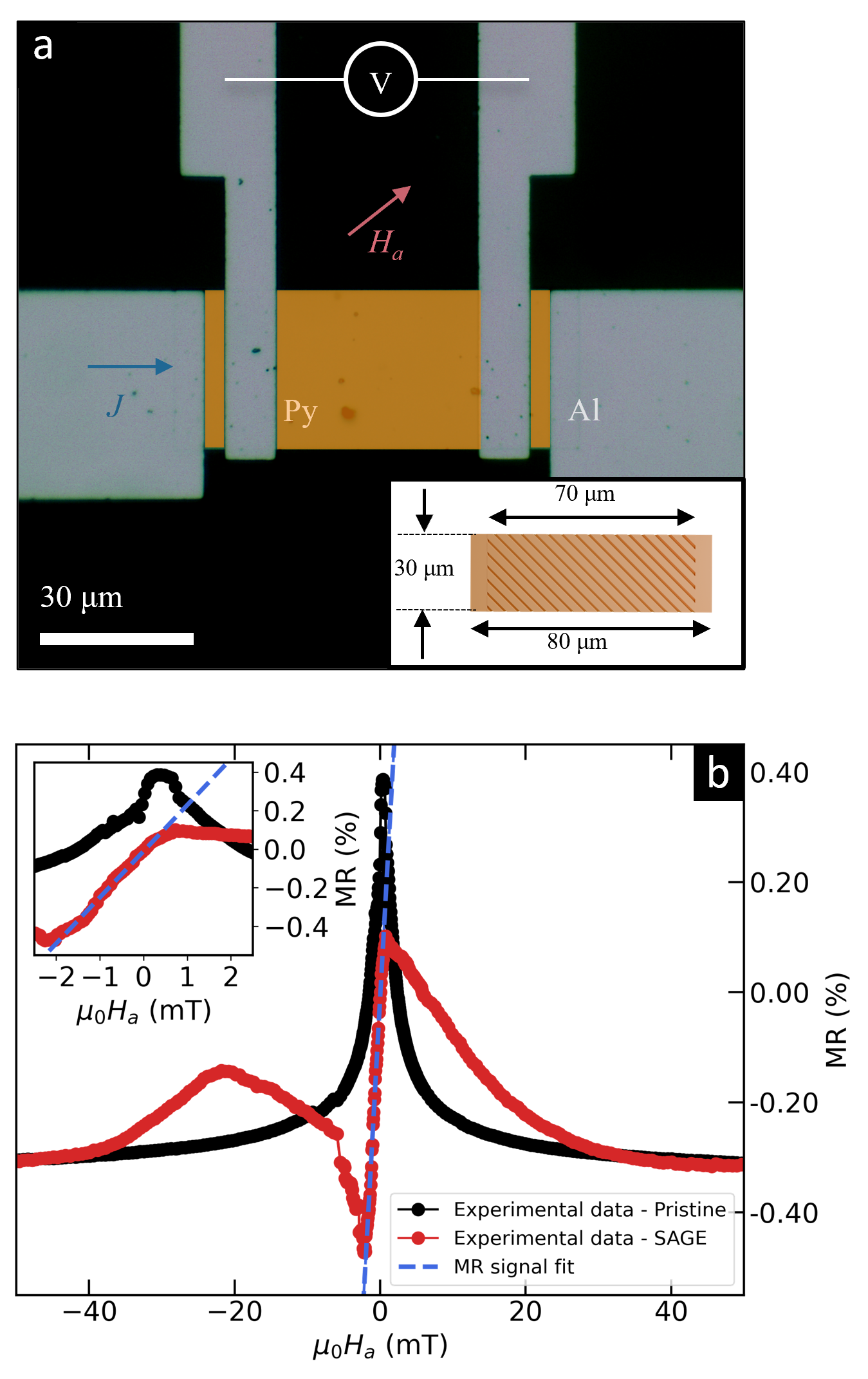}
    \caption{(a) False-colored optical microscopy image of the pristine AMR sensor, illustrating the directions of current density (\(J\), blue arrow) and applied magnetic field (\(H_a\), red arrow). The inset presents a schematic of the corrugated Py stripe with an angle of 45$^\circ$ between the EA and the current direction. The applied in-plane magnetic field is perpendicular to the EA. The SAGE-patterned region spans approximately 90\% of the stripe. (b) Comparison of the AMR response ($\mathrm{MR} = \Delta R/R_0$) as a function of the applied magnetic field for the corrugated Py stripe sensor (red) and the pristine Py sensor (black). For clarity, the pristine data have been vertically offset by 0.27\% to enhance visual distinction. The blue line represents a fit of the signal near its working point. The inset shows a magnified view of the magnetic field range over which the device operates. 
    } 
    \label{fig:AMR}
\end{figure}

AMR sensors based on soft FM stripes are typically optimized by engineering the current path using a barber-pole design \cite{1058886,8701473}. This layout allows for a nearly linear response to weak variations in the external magnetic field around 0 mT. Considering a device with a resistance $R=R_0$ when $\mu_0 H_a=0$ mT, the change in resistance $\Delta R=R-R_{0}$ is given by:

\begin{equation}
    \frac{\Delta R}{\Delta R_{m}} = \frac{H_a}{H_k} \sqrt{1 - \left( \frac{H_a}{H_k} \right)^2 }\simeq \dfrac{H_a}{H_k}
    \label{amrform}
\end{equation}

\noindent where $\Delta R_{\mathrm{m}} = R_{\mathrm{max}} - R_0$ denotes the maximum resistance change, $R_{\mathrm{max}}$ is the resistance when the magnetization is saturated along the current direction, and $H_k$ is the anisotropy field of the device. The final equality corresponds to the low-field approximation, valid when $H_a \ll H_k$. This expression predicts the AMR response of a FM stripe with a UMA lying at 45$^\circ$ with respect to the current flow, as a function of the external in-plane magnetic field applied perpendicularly to the UMA EA. 

Instead of redirecting the current flow, which reduces the effective working area and introduces current crowding effects~\cite{1058886, 8701473}, an alternative configuration consists of pinning the magnetization by means of a SAGE-induced UMA with the EA at 45$^\circ$ relative to the current direction, as shown in Fig.~\ref{fig:AMR}(a).

The efficiency of the proposed method is evidenced in Fig.~\ref{fig:AMR}(b), showing the response of an AMR sensor ($80 \times 30~\mu\mathrm{m}^2$) micro-engraved over nearly $90\%$ of its surface with a groove depth of $9.5~\mathrm{nm}$ and a periodicity of $300~\mathrm{nm}$. The AMR response to an external magnetic field applied at $90^{\circ}$ relative to the SAGE orientation is presented for the experimental SAGE-patterned device and compared with that of the pristine Py device under the same field orientation. The experimental data exhibit a trend consistent with the magnetoresistance signal given by Eq.~(\ref{amrform}), for an applied field range from -1.5 mT to 0.5 mT. From this fit, which represents the magnetoresistance (MR) behavior, the characteristic anisotropy field induced by the SAGE patterning was estimated to be $\mu_0 H_k \approx 6.5~\text{mT}$.

The sensitivity of an AMR sensor is defined as the rate of resistance variation induced by an external magnetic field change, divided by the total resistance $R$ of the device \cite{1058886}: 

\begin{equation}
S  = \frac{1}{\mu_0 R}\frac{dR}{dH_a}\simeq \dfrac{\Delta R_{m}}{\mu_0R_0 H_k}.
\end{equation}

In our SAGE AMR device, a sensitivity \(S\) as high as $2~\mathrm{m\Omega\,\Omega^{-1}\,mT^{-1}}$
 was measured over a linear range of \(-1.5\) to \(0.5~\mathrm{mT}\). 

A comparable performance (sensitivity of $6~\mathrm{m\Omega\,\Omega^{-1}\,mT^{-1}}$
 and a linear range from \(-2.5\) to \(2.5~\mathrm{mT}\)) has been demonstrated in an exchange-bias-based NiO/NiFe AMR device \cite{FMAFM} with a geometric aspect ratio of 20:1, and in which four barber-pole devices are arranged in a Wheatstone Bridge Configuration (WBC). For reference, highly-optimized commercial AMR sensors also on a WBC, such as the HMC1001~\cite{hmc1001}, report an enhanced sensitivity of $32~\mathrm{m\Omega\,\Omega^{-1}\,mT^{-1}}$
 within a narrower linear operating range of $-0.2$ to $0.2~\mathrm{mT}$. Note that achieving these performance levels requires significantly more complex fabrication processes and results in considerably less compact devices. In the SAGE barber-pole proposed in this work, the AMR signal’s sensitivity and symmetry can be further optimized by tuning the corrugation period, corrugation depth, and the aspect ratio (length : width) of the Py element \cite{QUYNH201698}, as well as by combining the components in a WBC.


\subsubsection{Zero-field magnonic device}

As a second illustrative application, a zero-field magnon waveguide is demonstrated. This idea has been intensively explored using various approaches, such as microcrystalline anisotropy \cite{Flajšman_2020}, non-collinear magnetic textures \cite{Wagner_2016, Yu_2021, Garcia-Sanchez_2015, Chang_2020}, and substrate-corrugation–induced anisotropy \cite{turcan_spin_2021, klima_zero-field_2024, montoncello_controlling_2021} to enable spin-wave propagation without the need of an external magnetic field. Similarly, SAGE can be used to propagate spin-waves in a FM medium without requiring any external field. 

\begin{figure*}
    \centering
    \includegraphics[width=0.95\linewidth]{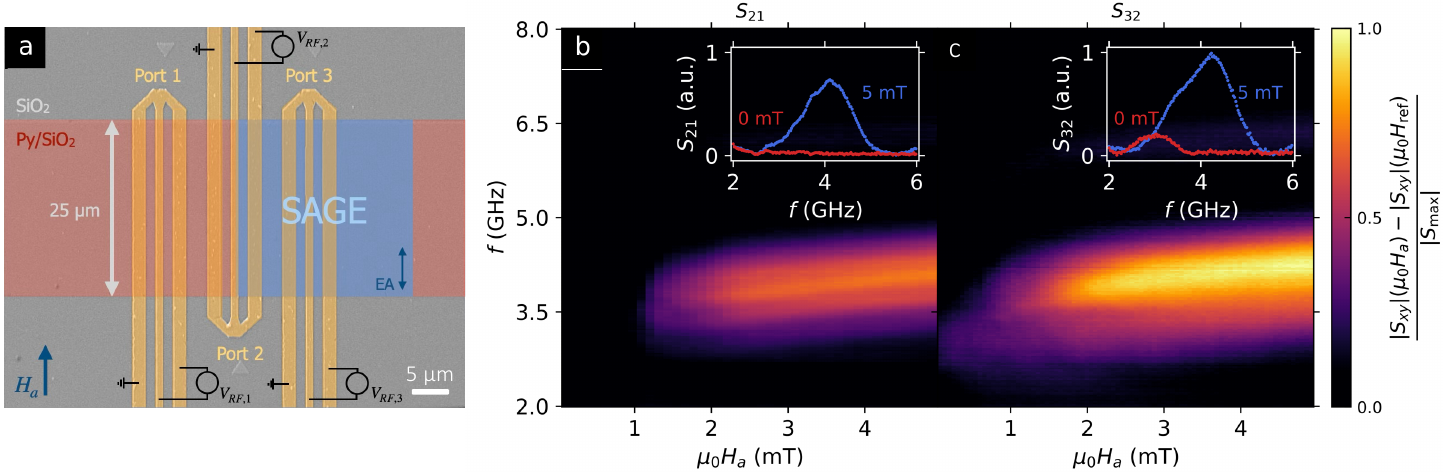}
    \caption{a) False-color scanning electron microscopy image of the device, comprising a ferromagnetic waveguide overlaid with three coplanar waveguide (CPW) antennas. The external magnetic field $H_a$ (blue arrow), oriented perpendicular to the spin-wave propagation direction (horizontal axis), enables the excitation of Damon–Eshbach spin waves.
    b) Normalized transmission parameter $S_{21}$ as a function of the applied magnetic field $H_a$ and excitation frequency $f$. The inset shows selected transmission spectra at 0 mT (red) and 5 mT (blue).
    c) Normalized transmission parameter $S_{32}$ as a function of $H_a$ and $f$. The inset shows selected transmission spectra at 0 mT (red) and 5 mT (blue).
    Each spectra is normalized by $S_\mathrm{max}=\mathrm{max}\{S_{32}(f, H)-S_{32}(f, H_\mathrm{ref})\}$}
    \label{fig:fig-psws}
\end{figure*}

To do so, a three-port device was fabricated, consisting of three coplanar waveguide (CPW) antennas coupled via a Py waveguide separated by an oxide spacer, as shown in Fig. \ref{fig:fig-psws}(a). Prior to deposition of the CPWs and the spacer, the right region of the waveguide (labeled SAGE and colored in blue in Fig. \ref{fig:fig-psws}(a)) was engraved to obtain 5 nm-deep grooves with a 400 nm spacing period and an EA orientation perpendicular to the Py stripe.

To control the magnetic state of the spin-wave waveguide, an external magnetic field was applied perpendicular to the radio-frequency field inductively generated by the CPWs, thereby exciting Surface Spin-Waves (SSW, also know as Damon–Eshbach modes) \cite{Damon_Eshbach_1961}. The transmission of the radio-frequency signal between the ports was characterized using a vector network analyzer under different applied magnetic fields $H_a$, swept from higher to lower values. The amplitudes of the insertion losses $S_{21}$ and $S_{32}$ are shown in Fig. \ref{fig:fig-psws}(b) and \ref{fig:fig-psws}(c). The non-magnetic background was evaluated at a reference field $H_{\mathrm{ref}}$ where no magnetic signal was present and subsequently subtracted to the each spectra. Afterward, each spectra, $S_{21}$ and $S_{32}$, was normalized by the highest $S_{32}(f, H)$ value (after background subtraction) denoted $S_{\mathrm{max}}$. 

The pristine side of the device (corresponding to $S_{21}$) exhibits the excitation of a spin wave with a wavevector determined from the CPW geometry \cite{delforge_investigation_2024}. A sharp reduction of the signal is observed at zero applied field. This behavior suggests that the necessary conditions for coherent SSW excitation, i.e. uniform magnetic domains oriented perpendicular to the waveguide axis, are disrupted \cite{Hubert_Schäfer_2011, Mahmoud_2020, Sadovnikov_2017}. Thus, in a pristine waveguide, a finite magnetizing field is always required to sustain SSW propagation. In contrast, in the SAGE region (see Fig. \ref{fig:fig-psws}(c)), coherent spin-wave propagation occurs even at zero applied field thanks to the induced anisotropy field favoring a single mono-domain oriented perpendicular to the waveguide. Direct comparison of the selected spectra at zero field (see insets) highlights the advantage of SAGE, enabling spin-wave transmission without an external magnetic field. The accompanying increase in amplitude and shift in resonance frequency, in agreement with earlier studies \cite{Flajšman_2020, turcan_spin_2021, Haldar_2017}, further confirm the effectiveness of this approach. Hence, SAGE represents a versatile platform for tuning spin-wave propagation.

\section{Conclusion}

The SAGE technique offers a robust and versatile route to tailoring magnetic anisotropy via surface corrugations. Its adaptability to diverse geometries, together with the programmable control afforded by SAGE-AFM, enables the design of arbitrary domain patterns and opens new opportunities for magnetic metamaterials and spatially modulated spintronic devices. While the present study has focused on straight, continuous grooves, the method can be extended to curved or discontinuous engravings, allowing local manipulation of the in-plane magnetic anisotropy $K_u(x,y)$ with micrometer-scale resolution. The main limitations remain its scalability to large areas and the relatively slow processing speed. Nonetheless, the applicability of this technique extends well beyond ferromagnetic systems, including superconductors, insulators, and non-magnetic films, thereby providing a powerful method for local anisotropy engineering and for probing novel physical phenomena.

\section{Methodology}
\subsection{Samples fabrication}
Pristine samples consist of disks and squares with diameters (or side lengths) ranging from 25 to 50~µm, fabricated by electron beam lithography followed by deposition of a 30~nm-thick Py thin film via electron beam evaporation and a subsequent lift-off process. SAGE was realized using an atomic force microscope (DriveAFM from Nanosurf) equipped with diamond tips (AD-40-AS) operated in contact mode, with a feedback loop ensuring constant interaction force between the tip and the sample. Prior to lithography, the cantilever was calibrated to determine the spring constant ($\,\mathrm{N\,m^{-1}}$) using the Sader method and the detection sensitivity ($\mathrm{nm\,V^{-1}}$) from force spectroscopy curve calibration, ensuring precise control of the applied force. Complex designs were generated by controlling the 2D displacement of the tip and the applied force on the sample surface via a Python script. SAGEs with varying parameters were investigated, with tip forces between 10 and 70~µN, periods between 150 and 550~nm, and either single or multiple passes, while the grooving speed was fixed at 4~$\mu\mathrm{m\,s^{-1}}$. After the SAGE process, nanomachining residues were removed by cleaning the samples in an IPA ultrasonic bath. The dimensions and homogeneity of the SAGE structures were characterized by AFM in tapping mode using a high aspect ratio tip (NW-HR-10). 

For the AMR sensor, electrical contacts, made of 150~nm-thick Al, were deposited by electron beam evaporation. Electrical measurements were performed using a Source Measure Unit Keithley 2400 in a four-point configuration with a DC current of 0.8~mA.

The spin-wave waveguide was fabricated and characterized following the procedure described in \cite{delforge_investigation_2024}. The SAGE parameters corresponding to each device used in this work are reported in Table \ref{tab:SAGE}.
\begin{table}[]
    \centering
    \begin{tabular}{|c|c|c|c|c|}
        \hline
        Device & SAGE Area (µm$^2$) & $\lambda$ (nm) & $w$ (nm) & $d$ (nm)\\
        \hline
         Squares & $25\times25$ & 150-550& 0-150& 0-15\\
         Disks & $\pi (25)^2$ & 300-600& 180-210 & 1 \\
         AMR sensor & $70\times30$ & 300 & 122 & 9\\
         SW waveguide & $25\times20$ &400 & 134 & 5 \\
         Chessboard & $50\times50$ & 400& 136 & 9\\
         \hline
    \end{tabular}
    \caption{Geometrical parameters of SAGE performed for this work.}
    \label{tab:SAGE}
\end{table}
\subsection{Experimental protocol}
Magnetic characterization was performed using a custom-built wide-field magneto-optical Kerr effect (MOKE) microscope in the longitudinal configuration, enabling direct imaging of in-plane magnetization states. The sample was illuminated by a white LED source through the microscope’s Köhler illumination, ensuring uniform, high-stability illumination  \cite{schafer_selective_2017}. A four-LED configuration allowed measurement of magnetization components both parallel and transverse to the applied field. The incident beam was linearly polarized by a high-extinction-ratio polarizer and focused onto the sample through a $50\times$ objective with numerical aperture 0.5. A quarter-wave plate compensator was inserted to correct the residual ellipticity \cite{shatalin}. The longitudinal Kerr signal was obtained by subtracting images recorded with opposite illumination directions, while summing such images enables sensitivity to out-of-plane magnetization. Reflected light passed through a beam splitter and an analyzer set $4^\circ$ from crossed, providing optimal contrast. Magnetization-encoded light was recorded using a Prime BSI Express CMOS camera. Post-processing included background subtraction method as described in \cite{fourneau_microscale_2023} to remove optical artifacts. Processed images were used to extract magnetization reversal loops under an externally applied in-plane magnetic field, which could be oriented along different directions to probe angular-dependent switching behavior.

\subsection{Micromagnetic simulations}
Micromagnetic simulations of the SAGE devices were performed using the open-source software MuMax$^3$\cite{vansteenkiste_design_2014}, which employs the finite-difference method to calculate the time- and space-dependent magnetization dynamics by solving the Landau–Lifshitz–Gilbert equation. The simulations were carried out assuming standard permalloy, with a saturation magnetization $M_s = 787\ \mathrm{kA\, m^{-1}}$, an exchange stiffness $
A_{\mathrm{ex}} = 1.3 \times 10^{-11}\ \mathrm{J\, m^{-1}}$, and no intrinsic magnetocrystalline anisotropy. The uniaxial magnetic anisotropy (UMA) induced by the SAGE process was modeled for 2 to 10 µm-wide square devices following two approaches: (a) as a homogeneous UMA with a fixed energy density $K_u$, or (b) by modifying the simulation box geometry to mimic ideal rectangular SAGE structures with fixed depth, width, and periodicity. For quantitative comparison with the 25 µm-wide experimental squares, the $H_c$ and $H_k$ values obtained for squares with different dimensions are extrapolated as explained in detail in the Supplementary Material. The discretization cell size was set to $\simeq 10$ nm along each spatial direction, except in simulations with SAGE-like geometries, where the out-of-plane cell dimension was reduced to match the smallest groove depth (i.e., 2.5 nm). In principle, this cell size exceeds the exchange length $\lambda \simeq 5$ nm, which could potentially reduce accuracy in describing domain wall characteristics. However, it was verified in a 8-µm-wide square device that both $H_c$ and $H_k$ remained nearly unchanged when using both 5 or 10 nm cell sizes. 
\\
\section{Acknowledgment}

The authors acknowledge financial support from Fonds de la Recherche Scientifique - FNRS and Flemish Research Foundation - FWO, under the grant Weave PDR T.0208.23-FNRS and G0D7723N-FWO, and by COST (European Cooperation in Science and Technology) [www.cost.eu] through COST Action SUPERQUMAP (CA 21144). K.T. acknowledges support from the KU Leuven C14/24/110 C1-program. C. D. and N. L. acknowledge support from FRS-FNRS (Research Fellowships FRIA). CzechNanoLab project LM2023051 funded by MEYS CR is gratefully acknowledged for the financial support of the measurements at CEITEC Nano Research Infrastructure. C.D. gratefully acknowledges insightful discussions with Michal Urb\'{a}nek and Ond\v{r}ej Wojewoda.The work of E.F. has
been financially supported by the FWO and F.R.S.-FNRS under the Excellence of Science (EOS) project O.0028.22.

\bibliographystyle{apsrev4-1} 
\bibliography{bib.bib}
\end{document}


\title{Supplementary Material: Engineering Magnetic Anisotropy in Permalloy Films via Atomic Force Nanolithography}

\author{Abhishek Naik} 
\affiliation{Experimental Physics of Nanostructured Materials,  Department of Physics, Universit\'{e} de Li\`{e}ge, B-4000 Sart Tilman, Belgium}

\author{Cyril Delforge}
\affiliation{Experimental Physics of Nanostructured Materials,  Department of Physics, Universit\'{e} de Li\`{e}ge, B-4000 Sart Tilman, Belgium}

\author{Nicolas Lejeune}
\affiliation{Experimental Physics of Nanostructured Materials,  Department of Physics, Universit\'{e} de Li\`{e}ge, B-4000 Sart Tilman, Belgium}

\author{Daniel Stoffels}
\affiliation{Experimental Physics of Nanostructured Materials,  Department of Physics, Universit\'{e} de Li\`{e}ge, B-4000 Sart Tilman, Belgium}

\author{Joris Van de Vondel}
\affiliation{Quantum Solid-State Physics, Department of Physics and Astronomy, KU Leuven, Celestijnenlaan 200D, Leuven, B-3001, Belgium}

\author{Kristiaan Temst}
\affiliation{Quantum Solid-State Physics, Department of Physics and Astronomy, KU Leuven, Celestijnenlaan 200D, Leuven, B-3001, Belgium}
\affiliation{Imec, Kapeldreef 75, 3001 Leuven, Belgium}

\author{Alejandro V. Silhanek} \email{asilhanek@uliege.be}
\affiliation{Experimental Physics of Nanostructured Materials,  Department of Physics, Universit\'{e} de Li\`{e}ge, B-4000 Sart Tilman, Belgium}

\author{Emile Fourneau} 
\affiliation{Experimental Physics of Nanostructured Materials,  Department of Physics, Universit\'{e} de Li\`{e}ge, B-4000 Sart Tilman, Belgium}

\date{\today}


\maketitle

    
\subsection{AFM Nanomachining} 

The interdependence between groove depth ($d$) and width ($w$) under varying applied normal forces and scan repetitions is presented in Fig.~\ref{fig:figSM0}(a). The data reveal a nearly linear correlation between $d$ and $w$, with an average ratio of $d/w \approx 0.1$. As shown in Fig.~\ref{fig:figSM0}(b), both groove depth and width increase progressively with the number of scans and with the applied normal force, respectively. The slope of these curves is expected to depend on the AFM tip displacement speed, as noted in previous reports\cite{phdthesis}, where higher engraving speeds were shown to decrease the groove depth while leaving the groove width largely unaffected. However the impact of the engraving tip velocity was not investigated in this work.



\begin{figure}[h!]
    \centering
    \includegraphics[width=0.8\linewidth]{Fig_SM1.png}
    \caption{(a) Correlation between the average corrugation width and depth in SAGE-patterned devices. The blue dashed line represents a linear fit to the data. (b) Variation of groove depth ($d$) and width ($w$) as a function of the number of scans, with the blue and red dashed lines indicating the corresponding linear fits.}

    \label{fig:figSM0}
\end{figure}

\subsection{Schl\"{o}mann's model}

The Schlömann model was used to estimate the anisotropy field ($H_k$) in square devices by approximating the surface corrugation as either perfectly triangular or rectangular grooves, as illustrated in the inset of Fig.~\ref{fig:figSM1}(a). The results, shown in Fig.~\ref{fig:figSM1} (red and blue lines), were compared with the $H_k$ values (black dots) calculated from the roughness measured by AFM imaging. This comparison shows that the experimentally observed anisotropy field aligns more closely with the predictions based on a rectangular-groove profile. The green-shaded region in the inset of Fig.~\ref{fig:figSM1}(a), corresponding to residual material generated during the SAGE process, was excluded from the roughness analysis since the Schlömann model requires a clean and well-defined surface modulation. Including these residues in the roughness calculation leads to an overestimation of the corrugation amplitude and thus produces unreliable $H_k$ values.

\begin{figure}[h!]
    \centering
    \includegraphics[width=0.9\linewidth]{supp-schloman.png}
   \caption{(a) Estimation of the anisotropy field $H_k$ using Schlömann's model for ideal rectangular (red line) and triangular (blue line) groove profiles, calculated for a fixed periodicity and varying groove depth in SAGE-patterned square devices. The inset shows a zoomed-in AFM line profile together with the rectangular (red dashed) and triangular (blue dashed) groove shape approximations used in the model. (b) Predicted $H_k$ values for grooves of fixed depth and varying period. The black dots correspond to model values obtained using the roughness parameters extracted from AFM measurements.}
    \label{fig:figSM1}
\end{figure}

\subsection{Micromagnetic simulations extrapolation toward 25 µm-large squares}

To bypass the computation memory limitation and simulate the uniaxial anisotropy in the 25 µm-large square devices, micromagnetic simulations were conducted on squares with edge lengths of 7--10~$\mu$m, using an anisotropy energy density constant, $K_u$, ranging from 0 to 12~kJ/m$^3$ and assumed to be homogeneous all over the device. The resulting anisotropy fields ($H_k$) and coercive fields ($H_c$) were extracted and extrapolated to 25~$\mu$m squares. The size-dependent extrapolation fits for different values of $K_u$ are plotted in Fig.~\ref{fig:figSM2}(a) and (b), respectively for $H_k$ and $H_c$. The extrapolated trends of $H_k$ and $H_c$ as functions of $K_u$ for the 25~$\mu$m squares are presented in Fig.~\ref{fig:figSM2}(c). 
  
  \begin{figure} [h!]
    \centering
    \includegraphics[width=0.9\linewidth]{mumax edge1.png}
    \caption{Micromagnetic simulation results showing (a) coercive field ($H_c$) and (b) anisotropy field ($H_k$) for square devices of varying sizes with in-plane uniaxial anisotropy. (c) Extrapolated $H_c$ and $H_k$ as a function of anisotropy energy density $K_u$ (kJ/m$^3$).}
    \label{fig:figSM2}
\end{figure}

\subsection{Correlation between anisotropy energy density and SAGE depth}

 As described in the methodology section of the main manuscript, micromagnetic simulations incorporating the SAGE pattern were performed using two complementary approaches: (i) a uniaxial magnetic anisotropy with a uniform energy density $K_u$, and (ii) an explicit modification of the device upper surface to form perfect cuboidal grooves, with a fixed periodicity of 300~nm and varying depths from 0 to 10~nm. Comparison of the coercive field $H_c$ obtained from both approaches for square devices of identical dimensions allows quantification of the UMA induced by SAGE. A similar comparison was performed for cuboidal grooves with a fixed depth of 8~nm and varying periodicity from 150 to 550~nm. The results, reported in Fig.~\ref{fig:figSM3}, reveal a linear relationship between the SAGE depth and the uniaxial anisotropy energy density, with an approximate slope of $K_u/d \approx 0.001~\mathrm{kJ/m^3/nm}$, and a similar linear correlation between the SAGE period and the induced anisotropy, with an approximate slope of $K_u/d \approx -0.59~\mathrm{kJ/m^3/nm}$.

\begin{figure} [h!]
    \centering
    \includegraphics[width=0.9\linewidth]{dnp.png}
    \caption{Micromagnetic simulation results showing the correlation between (a) SAGE period and (b) SAGE depth  with the average UMA energy density $K_u$. The orange line indicates the linear fit of the data.}
    \label{fig:figSM3}
\end{figure}

\section{Additional MOKE characterization}

The angular dependence of the coercive field $H_c(\theta)$ and the remanent magnetization $M_r(\theta)$ of ferromagnetic disks (radius $r = 25$~µm) with a fixed groove depth of 1.5~nm is shown in Fig.~\ref{fig:figpolar}(a) and (b) for SAGE periods of 300~nm (orange dots) and 600~nm (blue dots), respectively. The characteristic two-fold angular variation observed in both $H_c$ and $M_r/M_s$ confirms the presence of uniaxial magnetic anisotropy, which is consistent with the findings reported in the main manuscript. Small deviations (less than $10^\circ$) from the SAGE pattern orientation are observed for both 300~nm and 600~nm disks. This shift is consistent with a residual uniaxial magnetic anisotropy in the pristine samples, with an easy axis $120^\circ$ away from the SAGE direction. Furthermore, Fig.~\ref{fig:figpolar}(a) indicates that disks with 300~nm periodicity exhibit a stronger UMA compared to those with 600~nm periodicity, demonstrating that increasing the groove period systematically reduces the effective anisotropy of the system.

\begin{figure}[h!]
    \centering
    \includegraphics[width=0.9\linewidth]{polar SM.png}
    \caption{Angular dependence of (a) $H_c$ and (b) $M_r/M_{sat}$ for SAGE-patterned Py disks with 300~nm (orange) and 600~nm (blue) periodicity. Lines serve as a guide to the eye. Experimental data obtained via L-MOKE.}

    \label{fig:figpolar}
\end{figure}

\bibliography{bib.bib}